\documentclass[aps,prd,twocolumn,showpacs]{revtex4-1}

\usepackage{wallpaper}
\usepackage{bm}
\usepackage{amsmath}
\usepackage[colorlinks,linkcolor=blue,anchorcolor=blue,citecolor=red]{hyperref}

\begin{document}

\title{Interface effects of strange quark matter with density dependent quark masses}
\author{Cheng-Jun~Xia$^{1}$}
\email{cjxia@itp.ac.cn}
\author{Guang-Xiong~Peng$^{2,3,4}$}
\email{gxpeng@ucas.ac.cn}
\author{Ting-Ting~Sun$^{5}$}
\email{ttsunphy@zzu.edu.cn}
\author{Wan-Lei~Guo$^{3}$}
\email{guowl@ihep.ac.cn}
\author{Ding-Hui~Lu$^{6}$}
\email{dhlu@zju.edu.cn}
\author{Prashanth Jaikumar$^{7}$}
\email{Prashanth.Jaikumar@csulb.edu}

\affiliation{$^{1}${School of Information Science and Engineering, Ningbo Institute of Technology, Zhejiang University, Ningbo 315100, China}
\\$^{2}${School of Physics, University of Chinese Academy of Sciences, Beijing 100049, China}
\\$^{3}${Institute of High Energy Physics, Chinese Academy of Sciences, P.O. Box 918, Beijing 100049, China}
\\$^{4}${Synergetic Innovation Center for Quantum Effects and Application, Hunan Normal University, Changsha 410081, China}
\\$^{5}${School of Physics and Engineering and Henan Key Laboratory of Ion Beam Bioengineering, Zhengzhou University, Zhengzhou 450001, China}
\\$^{6}${Department of Physics, Zhejiang University, Hangzhou 310027, China}
\\$^{7}${Department of Physics and Astronomy, California State University Long Beach, 1250 Bellflower Blvd., Long Beach, California 90840, USA}}

\date{\today}

\begin{abstract}
We study the interface effects in strangelets adopting mean-field approximation (MFA). Based on an equivparticle model,
the linear confinement and leading-order perturbative interactions are included with density-dependent quark masses.
By increasing the confinement strength, the surface tension and curvature term of strange quark matter (SQM) become larger,
while the perturbative interaction does the opposite. For those parameters constrained according to the 2$M_\odot$ strange
star, the surface tension is $\sim$2.4 MeV/fm${}^2$, while unstable SQM indicates a slightly larger surface tension.
The obtained results are then compared with those predicted by the multiple reflection expansion (MRE) method.
In contrast to the bag model case, it is found that MRE method overestimates the surface tension and underestimates the
curvature term. To reproduce our results, the density of states in the MRE approach should be modified by proper damping
factors.
\end{abstract}


\maketitle

\section{\label{sec:intro}Introduction}

As the density of nuclear matter increases, a deconfinement phase transition is expected to take place and
form strange quark matter (SQM), which is comprised of approximately equal numbers of $u$, $d$, $s$ quarks.
Based on various quark models, it has been long suspected that SQM is the true ground state of strongly
interacting system~\cite{Bodmer1971_PRD4-1601, Witten1984_PRD30-272}, where stable lumps of SQM may exist
in our universe, e.g., strangelets~\cite{Farhi1984_PRD30-2379, Berger1987_PRC35-213, Gilson1993_PRL71-332,
Peng2006_PLB633-314}, nuclearites~\cite{Rujula1984_Nature312-734, Lowder1991_NPB24-177}, meteorlike compact
ultradense objects~\cite{Rafelski2013_PRL110-111102}, and strange stars~\cite{Itoh1970_PTP44-291,
Alcock1986_ApJ310-261, Haensel1986_AA160-121}. Nevertheless, if the dynamical chiral symmetry breaking
is considered, SQM may be unstable~\cite{Buballa1999_PLB457-261, Klahn2015_ApJ810-134}. In such cases,
it can only exist in extreme conditions, e.g., in the centre of compact stars~\cite{Weber2005_PPNP54-193,
Maruyama2007_PRD76-123015, Peng2008_PRC77-065807, Shao2013_PRD87-096012, Klahn2013_PRD88-085001,
Zhao2015_PRD92-054012, Li2015_PRC91-035803} and heavy-ion collisions~\cite{Greiner1987_PRL58-1825,
Greiner1991_PRD44-3517}. The properties of those SQM objects, the structures of SQM inside compact stars,
and the processes of quark-hadron transition, are sensitive to the interface effects, where the energy
contribution is often taken into account with a surface tension $\sigma$.

If SQM is absolutely stable, the quark-vacuum interface is crucial to the properties of SQM objects.
For example, adopting the bag model with bag constant $B$, it was shown that for $\sigma^{1/3}\approx B^{1/4}$
the surface effects could destabilize a small strangelet substantially~\cite{Farhi1984_PRD30-2379}.
A mass formula for strangelets was later derived where the minimum baryon number for metastable
strangelets increases linearly with $\sigma^3$~\cite{Berger1987_PRC35-213, Berger1989_PRD40-2128}.
Adopting a reasonable surface tension and including Coulomb interactions, it was found that large
strangelets are likely stable against fission~\cite{Heiselberg1993_PRD48-1418}. Meanwhile,
if $\sigma$ is small enough, there are strangelets at certain size more stable than
others~\cite{Alford2006_PRD73-114016}, where strange stars' surfaces may fragment into crystalline
crusts made of strangelets and electrons~\cite{Jaikumar2006_PRL96-041101}, or even form low-mass
large-radius strangelet dwarfs~\cite{Alford2012_JPG39-065201}. Beside the surface tension,
the curvature contribution also play important roles in small strangelets, where the multiple
reflection expansion (MRE) method was developed~\cite{Madsen1993_PRL70-391, Madsen1993_PRD47-5156,
Madsen1994_PRD50-3328}. The effects of electron-positron pair creation on the surface was found to
be crucial for the maximum net charge an object can carry~\cite{Madsen2008_PRL100-151102}. Since
the wavefunctions of quarks approach to zero on the quark-vacuum interface, the effects of quark
depletion was shown to be important for the properties of SQM objects~\cite{Xia2016_SciBull61-172,
Xia2016_SciSinPMA46-012021_E, Xia2016_PRD93-085025, Xia2017_JPCS861-012022, Xia2017_NPB916-669}.

If SQM is unstable, it may coexists with hadronic matter (HM) in compact stars, where the structure of
the quark-hadron mixed phase (MP) is affected by the quark-hadron interface~\cite{Maruyama2007_PRD76-123015}.
For a vanishing surface tension, the MP consists of point-like HM and SQM, which is in accordance with
the Glendenning construction~\cite{Glendenning2000}. For larger $\sigma$, due to the relocation of charged
particles on the quark-hadron interface, the geometrical structures such as droplet, slab, tube, and bubble
become stable~\cite{Heiselberg1993_PRL70-1355, Voskresensky2002_PLB541-93, Tatsumi2003_NPA718-359,
Voskresensky2003_NPA723-291, Endo2005_NPA749-333, Maruyama2007_PRD76-123015, Yasutake2012_PRD86-101302}.
Those geometrical structures get larger as we increase $\sigma$ and eventually the quark-hadron
interface becomes planar, which is consistent with those obtained with Maxwell construction.
The nonuniform distribution of SQM and HM could have important consequences on the properties of
compact stars, where the hyperon number fraction is suppressed and the neutrino opacity
enhanced~\cite{Reddy2000_PLB475-1}.

The transition between SQM and HM is also sensitive to the interface effects. For the creation of
SQM in neutron stars~\cite{Bombaci2016_EPJA52-58, Lugones2016_EPJA52-53}, core-collapse
supernova~\cite{Mintz2010_PRD81-123012, Mintz2010_JPG37-094066}, and heavy-ion
collisions~\cite{Toro2006_NPA775_102-126, Fogaca2016_PRC93-055204}, it was show that the surface
tension plays a crucial role, where larger $\sigma$ disfavors or inhibits quark matter nucleation
in HM. The survival of SQM objects in a heated environment is also sensitive to the value
of $\sigma$, where SQM objects may not have survived the evaporation or boiling process in the early
Universe~\cite{Alcock1985_PRD32-1273, Alcock1989_PRD39-1233, Lugones2004_PRD69-063509, Li2015_AP62-115}.

Despite the crucial importance, the interface effects of SQM are still poorly known. And even the
surface tension is not very well constrained. For vanishing chemical potentials, the surface tension
can be evaluated with lattice QCD, e.g., in Refs.~\cite{Huang1990_PRD42-2864, Huang1991_PRD43-2056,
Alves1992_PRD46-3678, Brower1992_PRD46-2703, Forcrand2005_NPB140-647, Forcrand2005_PRD72-114501}.
However, for finite chemical potentials, these calculations were haunted by the sign problem.
The surface tension can then only be estimated with effective models. For example, based on the linear
sigma model~\cite{Palhares2010_PRD82-125018, Pinto2012_PRC86-025203, Kroff2015_PRD91-025017},
Nambu-Jona-Lasinio model~\cite{Garcia2013_PRC88-025207, Ke2014_PRD89-074041}, three-flavor
Polyakov-quark-meson model~\cite{Mintz2013_PRD87-036004}, and Dyson-Schwinger equation
approach~\cite{Gao2016_PRD94-094030}, small values were obtained for the surface tension, i.e.,
$\sigma= 5 \sim 30\ \mathrm{MeV/fm}^{2}$. Adopting the quasiparticle model, Wen et al. predicted
$\sigma= 30 \sim 70\ \mathrm{MeV/fm}^{2}$ for the quark-vacuum interface~\cite{Wen2010_PRC82-025809}.
Based on Nambu-Jona-Lasinio model and adopting the MRE method, Lugones et al. obtained larger surface
tensions with $\sigma= 145 \sim 165\ \mathrm{MeV/fm}^{2}$~\cite{Lugones2013_PRC88-045803}.
For magnetized SQM, it was found that the surface tension has a different value in the parallel and
transverse directions with respect to the magnetic field~\cite{Lugones2017_PRC95-015804}. For
color-flavor locked SQM, the surface tension may be even larger, e.g., $\sigma\approx 300\
\mathrm{MeV/fm}^{2}$~\cite{Alford2001_PRD64-074017}.

In this work we investigate the interface effects of SQM in the equivparticle model~\cite{Peng2000_PRC62-025801,
Wen2005_PRC72-015204, Wen2007_JPG34-1697, Xia2014_SCPMA57-1304, Chen2012_CPC36-947,
Chang2013_SCPMA56-1730, Xia2014_PRD89-105027, Chu2014_ApJ780-135, Hou2015_CPC39-015101,
Peng2016_NST27-98, Chu2017_PRD96-083019}, where both linear confinement and leading-order perturbative
interactions are included with density-dependent quark masses~\cite{Xia2014_PRD89-105027}.
In particular, we study the properties of strangelets adopting mean-field approximation (MFA).
The obtained results are then compared with those of the MRE approach that overestimates the
surface tension and underestimates the curvature term in strangelets. This can be fixed by
introducing a modification to the density of states. The paper is organized as follows.
In Sec.~\ref{sec:the_Lagrangian}, we present the Lagrangian density of the equivparticle model.
The MFA is introduced in Sec.~\ref{sec:the_SletSM} with the quark wavefunctions obtained by
solving Dirac equations. Further simplifications with MRE method are presented in
Sec.~\ref{sec:the_SletMRE}, where the surface and curvature contributions to the density of
states are introduced. The obtained results are presented in Sec.~\ref{sec:num}.
Our conclusion is given in Sec.~\ref{sec:con}.

\section{\label{sec:the}Theoretical framework}
\subsection{\label{sec:the_Lagrangian}Lagrangian density}
The Lagrangian density of the equivparticle model can be given as
\begin{equation}
\mathcal{L} =  \sum_{i=u,d,s} \bar{\Psi}_i \left[ i \gamma^\mu \partial_\mu - m_i(n_\mathrm{b}) - e q_i \gamma^\mu A_\mu \right]\Psi_i
             - \frac{1}{4} A_{\mu\nu}A^{\mu\nu},  \label{eq:Lgrg_all}
\end{equation}
where $\Psi_i$ represents the Dirac spinor of quark flavor $i$, $m_i(n_\mathrm{b})$ the mass, and $A_\mu$ the photon field with the field tensor
\begin{equation}
A_{\mu\nu} = \partial_\mu A_\nu - \partial_\nu A_\mu.
\end{equation}
In the equivparticle model, the strong interactions are considered with density-dependent quark masses and quarks are treated as quasi-free
particles. Taking into account both the linear confinement and leading-order perturbative interactions, the quark mass scaling is given
by~\cite{Xia2014_PRD89-105027}
\begin{equation}
  m_i(n_{\mathrm b})=m_{i0} + m_\mathrm{I}(n_{\mathrm b})=m_{i0}+\frac{D}{\sqrt[3]{n_\mathrm{b}}}+C\sqrt[3]{n_\mathrm{b}}.  \label{Eq:mnbC}
\end{equation}
Here $m_{i0}$ is the current mass of quark flavor $i$ with $m_{u0}=2.2$ MeV, $m_{d0}=4.7$ MeV, and $m_{s0}=96.0$ MeV~\cite{PDG2016_CPC40-100001}.
The confinement parameter $D$ is connected to the string tension $\sigma_0$, the chiral restoration density $\rho^*$, and the
sum of the vacuum chiral condensates $\sum_q\langle\bar{q}q\rangle_0$. Meanwhile, the perturbative strength parameter $C$ is
linked to the strong coupling constant $\alpha_\mathrm{s}$. The baryon number density is given by $n_\mathrm{b}= \sum_{i=u,d,s}n_i/3$
with the number density $n_i=\langle \bar{\Psi}_i \gamma^0 \Psi_i\rangle$. Adopting the mean-field and no-sea approximations,
the single particle Dirac equations for quarks and Klein-Gordon equation for photons are obtained via a variational procedure.
Note that electrons are neglected here since their contributions are comparatively small for strangelets with radii $R \lesssim 40$ fm.
For larger strangelets, however, one should not neglect them due to electron-positron pair creation~\cite{Madsen2008_PRL100-151102,
Xia2017_JPCS861-012022}.

One important aspect for density dependent models, being the dependence included either in the mass or in the coupling terms,
is the self-consistency of thermodynamics. There have been many efforts in the literature dealing with this problem in models similar
to the one used in the present work, e.g., Refs.~\cite{Brown1991_PRL66-2720, Wang2000_PRC62-015204, Peng2000_PRC62-025801,
Torres2013_EPL101-42003, Dexheimer2013_EPJC73-2569}. In principle, any effective models should meet the requirement of fundamental
thermodynamics, where all the quantities are derived accordingly. To show this explicitly, in Ref.~\cite{Xia2014_PRD89-105027} we have
proved necessary conditions for self-consistent thermodynamics, and it was shown that many thermodynamic treatments are inconsistent,
e.g., in Refs.~\cite{Chakrabarty1989_PLB229-112, Wang2000_PRC62-015204, Benvenuto1995_PRD51-1989}. Since here we have adopted a density
dependent mass scaling in Eq.~(\ref{Eq:mnbC}), in obtaining the equation of motion or other thermodynamic quantities, it is essential
that we include the density derivative terms of quark masses, which is discussed in Secs.~\ref{sec:the_SletSM} and~\ref{sec:the_SletMRE}.

\subsection{\label{sec:the_SletSM} Strangelets in MFA}
For spherically symmetric strangelets, the Dirac spinor of quarks can be expanded as
\begin{equation}
 \psi_{n\kappa m}({\bm r}) =\frac{1}{r}
 \left(\begin{array}{c}
   iG_{n\kappa}(r) \\
    F_{n\kappa}(r) {\bm\sigma}\cdot{\hat{\bm r}} \\
 \end{array}\right) Y_{jm}^l(\theta,\phi)\:,
\label{EQ:RWF}
\end{equation}
with $G_{n\kappa}(r)/r$ and $F_{n\kappa}(r)/r$ being the radial wave functions for the upper and lower components, while $Y_{jm}^l(\theta,\phi)$
is the spinor spherical harmonics. The quantum number $\kappa$ is defined by the angular momenta $(l,j)$ as $\kappa=(-1)^{j+l+1/2}(j+1/2)$.

Then the Dirac equation for the radial wave functions is obtained as
\begin{equation}
 \left(\begin{array}{cc}
  V_i + V_S                                                   & {\displaystyle -\frac{\mbox{d}}{\mbox{d}r} + \frac{\kappa}{r}}\\
  {\displaystyle \frac{\mbox{d}}{\mbox{d}r}+\frac{\kappa}{r}} & V_i - V_S - 2 m_{i0}                      \\
 \end{array}\right)
 \left(\begin{array}{c}
  G_{n\kappa} \\
  F_{n\kappa} \\
 \end{array}\right)
 = \varepsilon_{n\kappa}
 \left(\begin{array}{c}
  G_{n\kappa} \\
  F_{n\kappa} \\
 \end{array}\right) \:,
\label{Eq:RDirac}
\end{equation}
with the single particle energy $\varepsilon_{n\kappa}$, the mean field scalar and vector potentials
\begin{eqnarray}
 V_S &=& m_\mathrm{I}(n_\mathrm{b}), \label{Eq:Vs}\\
 V_i &=& \frac{1}{3}\frac{\mbox{d} m_\mathrm{I}}{\mbox{d} n_\mathrm{b}}\sum_{i=u,d,s}  n_i^\mathrm{s} + e q_i A_0. \label{Eq:Vv}
\end{eqnarray}
Note that in deriving the vector potentials based on variational method, we have obtained the density derivative terms
of quark masses since $n_\mathrm{b}= \sum_{i=u,d,s} \langle \bar{\Psi}_i \gamma^0 \Psi_i\rangle/3$, which is similar to introducing
the ``rearrangement" term in relativistic-mean-field models due to the density dependent coupling constants~\cite{Lenske1995_PLB345-355}.
Since the vector potentials for different quarks share a common term, we define $V_V = \frac{1}{3}\frac{\mbox{d} m_\mathrm{I}}{\mbox{d} n_\mathrm{b}}
\sum_{i=u,d,s}  n_i^\mathrm{s}$ and Eq.~(\ref{Eq:Vv}) becomes $V_i = V_V + e q_i A_0$.

The Klein-Gordon equation for photons is given by
\begin{equation}
- \nabla^2 A_0 = e n_\mathrm{ch}. \label{Eq:K-G}
\end{equation}
where $n_\mathrm{ch} = \sum_i q_i n_i$ is the charge density with $q_u = 2/3, q_d = -1/3$, and $q_s = -1/3$.

For given radial wave functions, the scalar and vector densities for quarks can be determined by
\begin{subequations}
\begin{eqnarray}
 n_i^\mathrm{s}(r) &=& \frac{1}{4\pi r^2}\sum_{k=1}^{N_i}
 \left[|G_{k i}(r)|^2-|F_{k i}(r)|^2\right] \:,
\\
 n_i(r) &=& \frac{1}{4\pi r^2}\sum_{k=1}^{N_i}
 \left[|G_{k i}(r)|^2+|F_{k i}(r)|^2\right] \:,
\end{eqnarray}%
\label{Eq:Density}%
\end{subequations}%
where the quark numbers $N_i\ (i=u,d,s)$ are obtained by integrating the density $n_i(r)$ in coordinate space as
\begin{equation}
 N_i = \int 4\pi r^2 n_i(r) \mbox{d}r. \label{Eq:axi}
\end{equation}

Finally, the total mass of a strangelet can be obtained with
\begin{eqnarray}
M &=& \sum_{i=u,d,s}\sum_{k=1}^{N_i} (\varepsilon_{ki} + m_{i0}) - \int 12\pi r^2 n_\mathrm{b}(r) V_V(r) \mbox{d}r \nonumber \\
  &\mathrm{}& - \int 2\pi r^2 n_\mathrm{ch}(r) e A_0(r) \mbox{d}r.
\label{Eq:M}
\end{eqnarray}

For given $C$ and $D$, we solved the Dirac Eq.~(\ref{Eq:RDirac}), mean field potentials Eq.~(\ref{Eq:Vs}) and Eq.~(\ref{Eq:Vv}),
Klein-Gordon Eq.~(\ref{Eq:K-G}), and densities Eq.~(\ref{Eq:Density}) inside a box by iteration in coordinate space with the
grid width $0.005$ fm. The box size $R$ varies with quark numbers and is fixed at vanishing densities.

\subsection{\label{sec:the_SletMRE} Strangelets with MRE method}
To investigate strangelets in MFA in Sec.~\ref{sec:the_SletSM} is straightforward, but numerically demanding. Instead of
solving the wavefunctions, we can simplify our calculation further by adopting the MRE method~\cite{Berger1987_PRC35-213,
Madsen1993_PRL70-391, Madsen1993_PRD47-5156, Madsen1994_PRD50-3328}, where the average interface effects are
treated with a modification to the density of states including both surface and curvature contributions, i.e.,
\begin{eqnarray}
N_i^{\prime}(p) &=& 6\left[ \frac{p^2 v}{2\pi^2} + f_s \left(\frac{p}{m_i}\right) p s + f_c \left(\frac{p}{m_i}\right) c \right], \label{Eq:niexpr} \\
f_s(x) &=& - \frac{\eta_\mathrm{s}}{4\pi^2}\arctan\left(\frac{1}{x}\right), \label{Eq:fs}\\
f_c(x) &=& \frac{\eta_\mathrm{c}}{12\pi^2}\left[1-\frac{3}{2}x \arctan\left(\frac{1}{x}\right)\right]. \label{Eq:fc}
\end{eqnarray}
To make the density of states more flexible to the fact that the surface tension and curvature contribution may be affected by
the medium effects, we introduce a surface strength factor $\eta_\mathrm{s}$ in Eq.~(\ref{Eq:fs}), and a curvature strength factor
$\eta_\mathrm{c}$ in Eq.~(\ref{Eq:fc}). Normally, they are equal to unity, as in the original literature~\cite{Farhi1984_PRD30-2379,
Madsen1993_PRD47-5156} where the strong quark interactions considered are mainly confinement by the bag constant. In the present
contact, we consider both confinement and perturbative interactions. We will therefore find that it is more appropriate to take
$\eta_\mathrm{s}\approx 0.3$ and $\eta_\mathrm{c}\approx 0.1$.

For a spherically symmetric system with radius $R$, the volume $v=4\pi R^3/3$, surface area $s=4\pi R^2$, and curvature $c=8\pi R$.
The number of quarks can then be obtained with
\begin{equation}
N_i = \int_0^{\nu_i} N_i'(p) \mbox{d}p, \label{Eq:Ni}
\end{equation}
where $\nu_i$ corresponds to the Fermi momentum of quark flavor $i$. The number of depleted quarks on a strangelet's surface can
be determined by subtracting the volume term, i.e., $N_i^\mathrm{surf} = N_i - {4\nu_i^3 R^3}/{3\pi}$.

In equivparicle model, however, it is more convenient to work in densities. Based on Eq.~(\ref{Eq:Ni}), the average quark number
density can be obtained with $n_i = \int_0^{\nu_i} N_i'(p) \mbox{d}p/V = 3 N_i(\nu_i)/4\pi R^3$, which gives
\begin{eqnarray}
n_i&=& \frac{\nu_i^3}{\pi^2} + \frac{9\eta_\mathrm{S} m_i^2}{4\pi^2R}
          \left[y_i^2\arctan(x_i)-x_i(\frac{\pi x_i}{2}+1)\right] \nonumber\\
&\mathrm{}& + \frac{9\eta_\mathrm{C} m_i}{4\pi^2R^2}
          \left[y_i^2\arctan(x_i)- x_i(\frac{\pi x_i}{2}-\frac{1}{3})\right].
\label{Eq:ni}
\end{eqnarray}
Here $x_i\equiv \nu_i/m_i$ and $y_i \equiv \sqrt{x_i^2+1}$. The average baryon number density adopted in Eq.~(\ref{Eq:mnbC}) is
then obtained with $n_\mathrm{b}= \sum_{i=u,d,s}n_i/3$.

Based on the dispersion relation $\epsilon_i=\sqrt{p^2+m_i^2}$, the mass of a strangelet is given by
\begin{equation}
M = \sum_{i=u,d,s} \int_0^{\nu_i} \sqrt{p^2+m_i(n_\mathrm{b})^2} N_i'(p) \mbox{d}p + M_\mathrm{ch}. \label{Eq:M_MRE}
\end{equation}
The first term includes the kinetic energy and strong interactions, while $M_\mathrm{ch}$ is the Coulomb energy with
\begin{equation}
M_\mathrm{ch}=\frac{8}{45}\pi^2 R^5 e^2 ({n_\mathrm{ch}^\mathrm{v}}^2+5 n_\mathrm{ch}^2). \label{Eq:Mch}
\end{equation}
Here $n_\mathrm{ch}^\mathrm{v} = \sum_i q_i \nu_i^3/\pi^2$ is the volume term of the total electric charge density $n_\mathrm{ch}=
\sum_i q_i n_i$. Note that the Debye screening effects are not included here, which is important to determine the critical size of
stable strangelets and reduces the charge-to-mass ratio of strangelets with $A\gtrsim 10^5$~\cite{Alford2006_PRD73-114016,
Xia2017_JPCS861-012022}. Based on the fundamental thermodynamic relations, one can obtain the chemical potential $\mu_i = \left.
\frac{\partial M}{\partial N_i}\right|_{V, \{N_k\neq i\}}$ and pressure $P = -\left.\frac{\partial M}{\partial V}\right|_{\{N_i\}}$.
It is worth mentioning that the density derivative terms of quark masses appear in the expressions of $\mu_i$ and
$P$~\cite{Xia2014_SCPMA57-1304}, which are essential to maintain the self-consistency of thermodynamics. For detailed derivations
and explanations, one can refer to Refs.~\cite{Peng2008_PRC77-065807, Xia2014_PRD89-105027, Xia2014_SCPMA57-1304}.

\section{\label{sec:num}Results and discussions}
For a strangelet with given total baryon number $A$, the particle numbers of $u$, $d$, $s$ quarks are fixed so that
its mass $M$ reaches minimum, i.e., fulfilling the $\beta$-stability condition $\mu_u = \mu_d = \mu_s$. When the
MRE method is adopted, the radius $R$ is determined according to the mechanic stability condition, i.e., $P = 0$.

Since its initial proposal~\cite{Xia2014_PRD89-105027}, the mass scaling in Eq.~(\ref{Eq:mnbC}) has been adopted and
examined in various occasions~\cite{Xia2014_PRD89-105027, Xia2015_CAA56-79, Hou2015_CPC39-015101, Peng2016_NST27-98,
Peng2018_JPSCP20-011022, Zhou2018_IJMPE27-1850037}. For example, in Ref.~\cite{Xia2014_PRD89-105027} we investigated the stability
window of SQM for the confinement parameter $D$ and perturbative strength parameter $C$, which is then confronted with
pulsar observation. On the one hand, if SQM is absolutely stable, a strange star obtained with $C\gtrsim 0.6$  can be more
massive than PSR J0348+0432~\cite{Antoniadis2013_Science340-6131}, where the parameter sets ($C$, $\sqrt{D}$):
(0.6, 133 MeV) and (0.7, 129 MeV) are found to be reasonable~\cite{Xia2014_PRD89-105027, Xia2015_CAA56-79, Peng2018_JPSCP20-011022}.
For strangelets, recently it was shown that the perturbative interaction increases their radii and masses~\cite{Zhou2018_IJMPE27-1850037}.
On the other hand, unstable SQM can only exist in the core of a hybrid star as mixed phases with hadronic matter,
where the hybrid star can be more massive than 2$M_\odot$ for $C = 0.7$, and $\sqrt{D} = 170$, 190 MeV~\cite{Li2015_PRC91-035803}.
In this work, to examine all possibilities, we adopt the parameter sets ($C$, $\sqrt{D}$): (0.4, 129 MeV),
(0.7, 129 MeV), and (0.7, 140 MeV). The properties of strangelets are then investigated adopting MFA in
Sec.~\ref{sec:the_SletSM} and MRE method in Sec.~\ref{sec:the_SletMRE}.

\begin{figure}
\includegraphics[width=\linewidth]{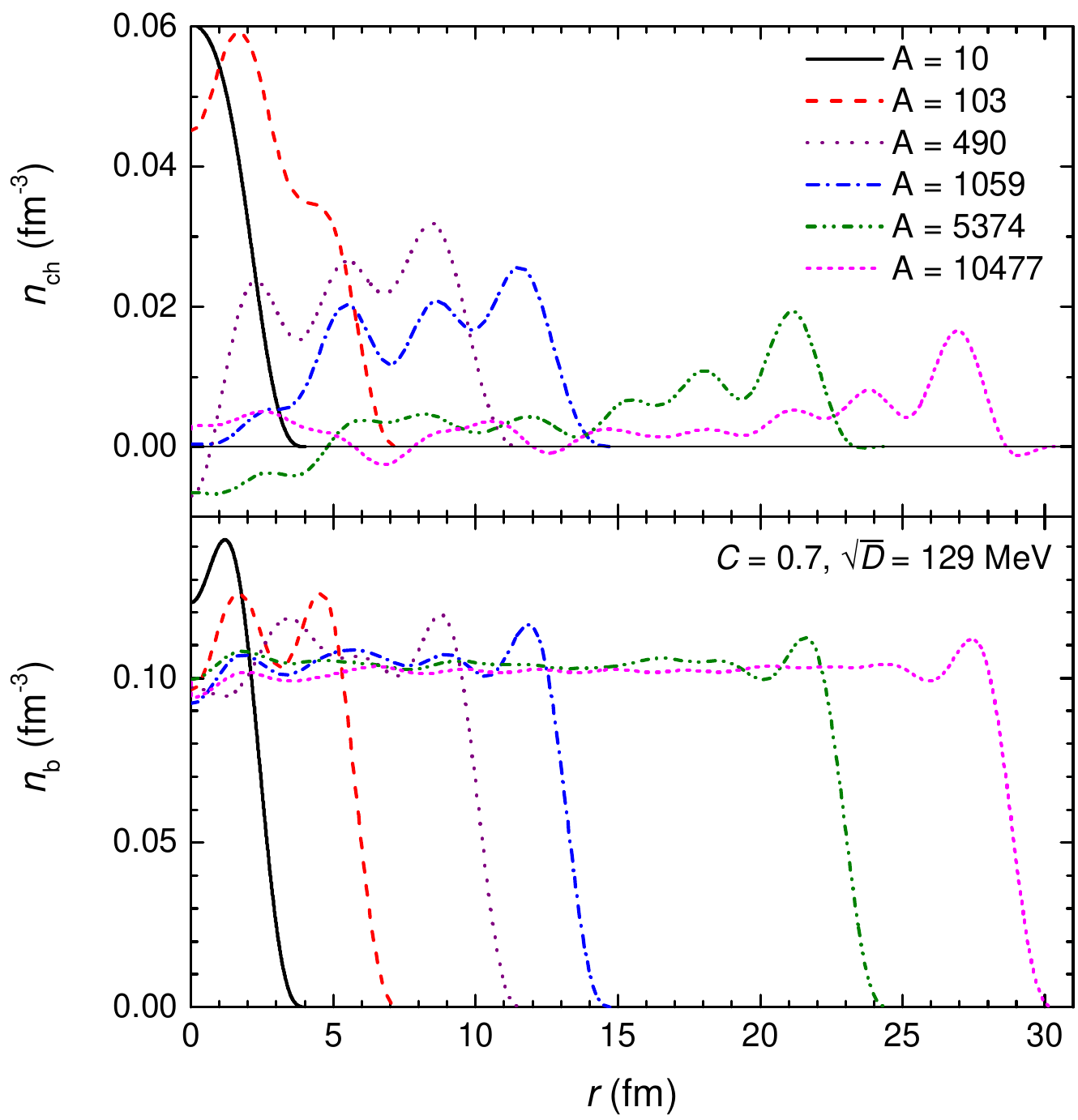}
\caption{\label{Fig:nBch_D129C07} The baryon and charge density distribution in a strangelet at various total baryon numbers.}
\end{figure}

In Fig.~\ref{Fig:nBch_D129C07} we present the baryon number density and charge density for strangelets, which was predicted
by taking $C = 0.7$ and $\sqrt{D} = 129$ MeV. As the baryon number $A$ increases, a strangelet becomes larger with a
smoother internal baryon number density distribution, and is approaching to the bulk value $n_0 = 0.099\ \mathrm{fm}^{-3}$.
Note that $n_0$ is smaller than nuclear saturation density $0.16\ \mathrm{fm}^{-3}$. This can be fixed if we include the
zero-point energy~\cite{DeGrand1975_PRD12-2060} and center-of-mass correction, which could shrink baryons. The surface
baryon number density distribution varies little with $A$, which will be discussed in more detail later. The charge
densities, however, varies with baryon number. At small $A$, internally a strangelet can be more positively charged.
For larger $A$, due to Coulomb repulsion, a strangelet can carry charge only on its surface and the surface structure
starts to converge. This was also predicted by the UDS model, where larger SQM objects indicate a constant surface charge
density~\cite{Xia2016_SciBull61-172, Xia2016_SciSinPMA46-012021_E, Xia2016_PRD93-085025, Xia2017_JPCS861-012022, Xia2017_NPB916-669}.

\begin{figure}
\includegraphics[width=\linewidth]{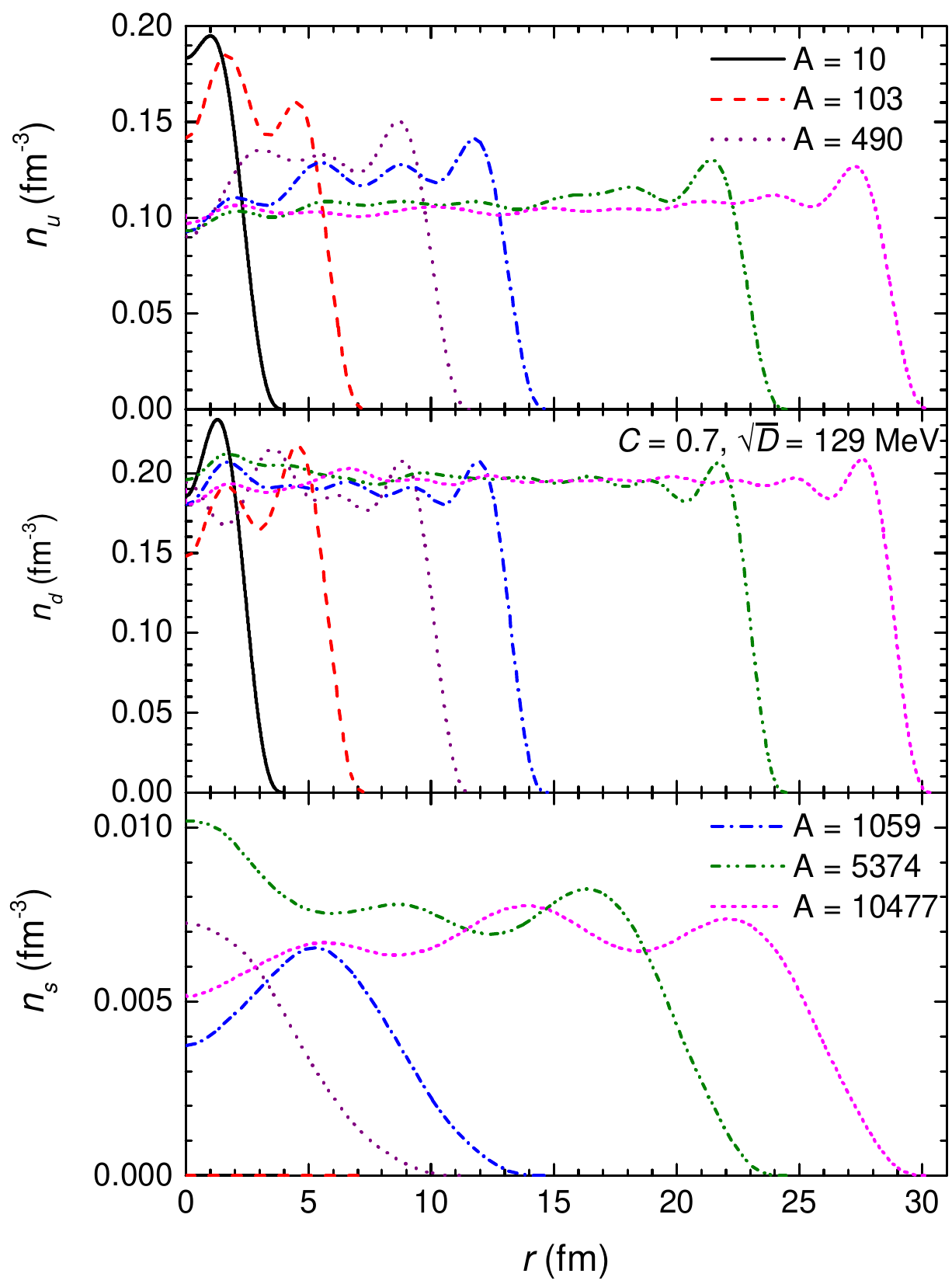}
\caption{\label{Fig:ni_D129C07} Density profiles of $u$-, $d$-, $s$-quarks for strangelets with various baryon numbers.}
\end{figure}

To give a more detailed picture on the internal structure of strangelets, in Fig.~\ref{Fig:ni_D129C07} we present the density
distributions of $u$-, $d$-, $s$-quarks corresponding to Fig.~\ref{Fig:nBch_D129C07}. It is found that the $d$-quark density
inside a strangelet varies little with baryon number and is close to the bulk density $n_d = 0.19\ \mathrm{fm}^{-3}$.
For small strangelets without strangeness ($A\lesssim200$), $u$-quark density inside a strangelet is close to the $d$-quark
density. Once $s$-quarks start to appear at $A\gtrsim 200$, as $A$ increases, the internal density for $u$-quarks is decreasing
with increasing $s$-quark density, which are approaching to their bulk values $n_u = 0.099\ \mathrm{fm}^{-3}$ and
$n_s = 0.0055\ \mathrm{fm}^{-3}$. Similar to Fig.~\ref{Fig:nBch_D129C07}, the surface density distributions for $u$-,
$d$-, $s$-quarks vary little with $A$. Comparing to $u$-, $d$-quarks, $s$-quarks are more diffused on a strangelet's surface.
This is because the strangelets carry much less $s$-quarks than $u$-, $d$-quarks, which is mainly caused by the large
$s$-quark mass.

\begin{figure}
\includegraphics[width=\linewidth]{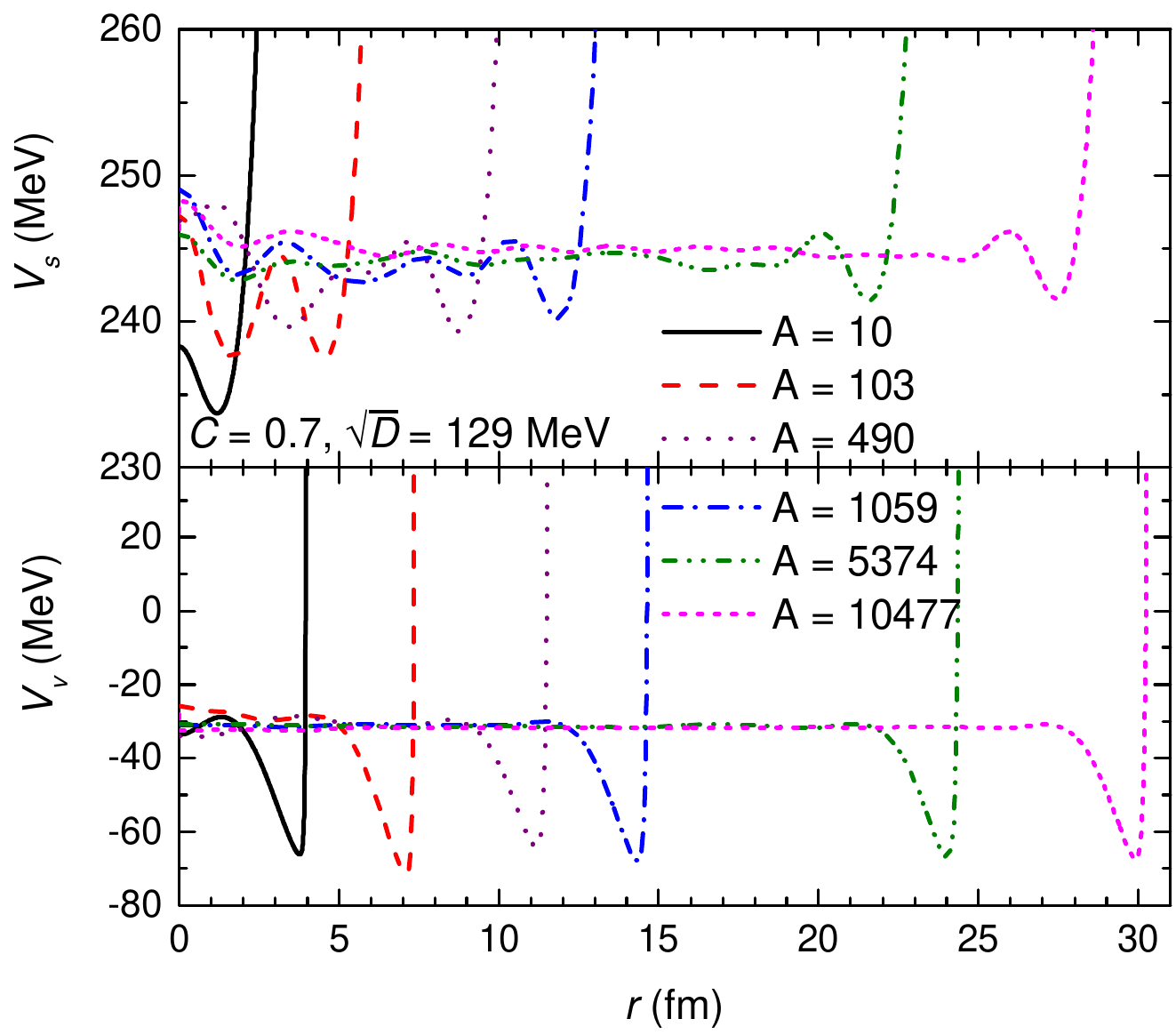}
\caption{\label{Fig:Vsv_D129C07} Scalar and vector potentials inside strangelets.}
\end{figure}

For given density distributions, the mean field potentials are obtained with Eq.~(\ref{Eq:Vs}) and Eq.~(\ref{Eq:Vv}).
In Fig.~\ref{Fig:Vsv_D129C07} we show the scalar and vector potentials corresponding to the densities in
Figs.~\ref{Fig:nBch_D129C07} and~\ref{Fig:ni_D129C07}. Note that only the common term $V_V$ in vector potentials
is presented in Fig.~\ref{Fig:Vsv_D129C07}. Since quark confinement is taken into account self-consistently in
our mass scaling Eq.~(\ref{Eq:mnbC}), the mean field potentials become infinitely large at the quark-vacuum interface.
As baryon number increases, the potential depth of $V_S$ increases while $V_V$ varies little.
Since the internal density distributions become smoother at larger $A$, the potentials become smoother as well.
Meanwhile, the potentials in the vicinity of quark-vacuum interfaces converge as one increases $A$.

\begin{figure}
\includegraphics[width=\linewidth]{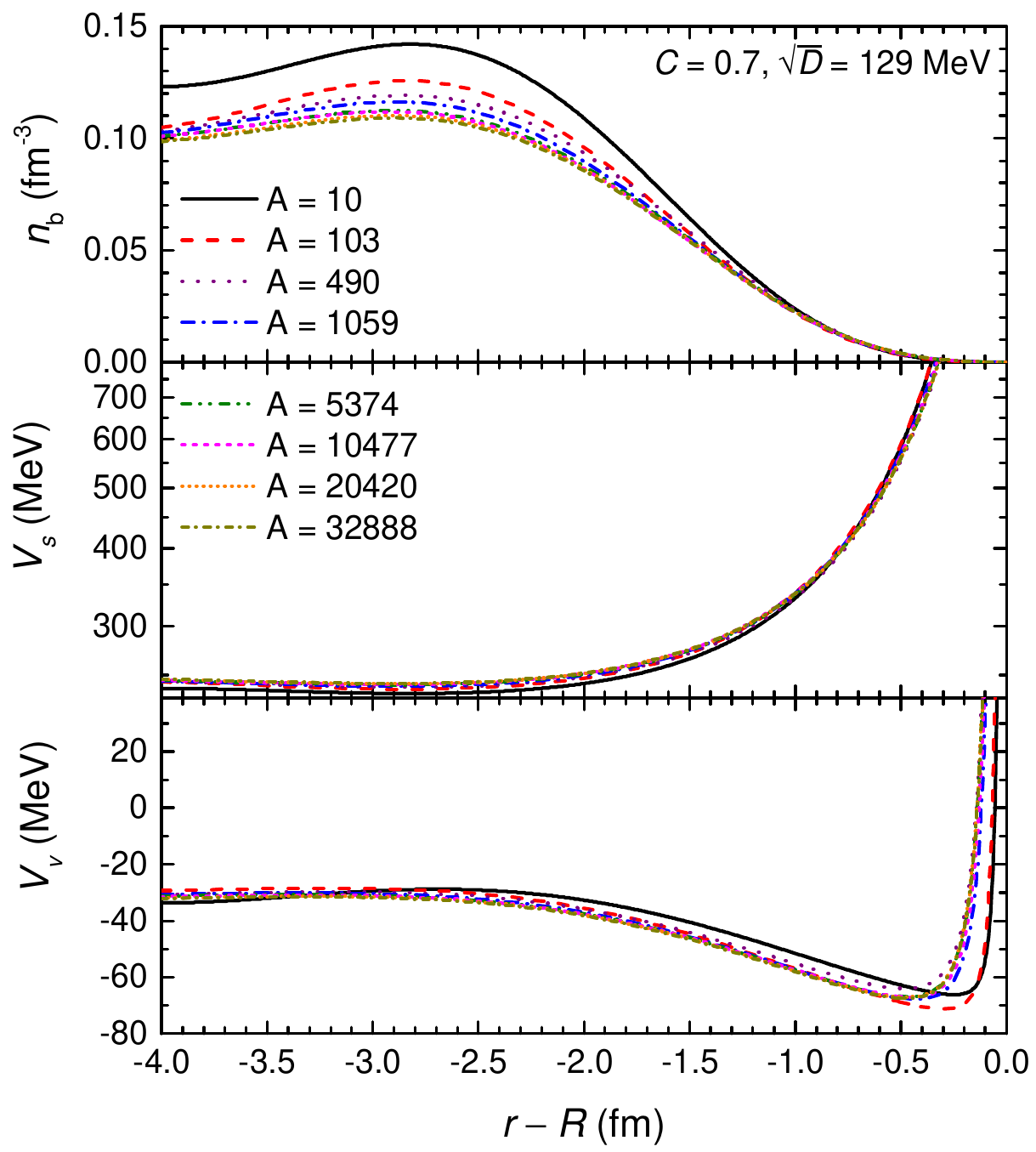}
\caption{\label{Fig:Surf_D129C07} Density and potential profiles on the surface of strangelets. The box size $R$ are fixed at vanishing densities.}
\end{figure}

To examine the surface structures of strangelets and their relevance to interface effects of SQM, in Fig.~\ref{Fig:Surf_D129C07}
we show the baryon number density, scalar and vector potentials in the vicinity of quark-vacuum interfaces. At around
2.8 fm beneath the surface, the density starts to drop and slowly approaches to zero on the surface. This is essentially
different from bag model predictions, where a sudden drop of density on the quark-vacuum interface is observed, e.g.,
in Ref.~\cite{Oertel2008_PRD77-074015}. The main reason is that equivparticle model reaches confinement with density
dependent quark masses while bag model introduces an infinite wall. Since the lattice calculation suggests that quark-quark
interaction is proportional to the distance~\cite{Belyaev1984_PLB136-273} instead of a wall, the surface density profiles
in Fig.~\ref{Fig:Surf_D129C07} are more reasonable. For smaller strangelets, larger internal densities (at $R-r\approx 2.8$ fm)
are obtained so that the density drops faster on the surface. In such cases, as indicated in Fig.~\ref{Fig:Surf_D129C07},
the potentials obtained with Eq.~(\ref{Eq:Vs}) and Eq.~(\ref{Eq:Vv}) vary more drastically with $r$ on the surface.
The effects observed in smaller strangelets can essentially be attributed to the curvature term, which has more contribution
on the properties of smaller strangelet. As one increases the baryon number $A$, the curvature term becomes insignificant
and the surface structures of strangelets start to converge and varies little for $A\gtrsim 10^5$.

\begin{figure}
\includegraphics[width=\linewidth]{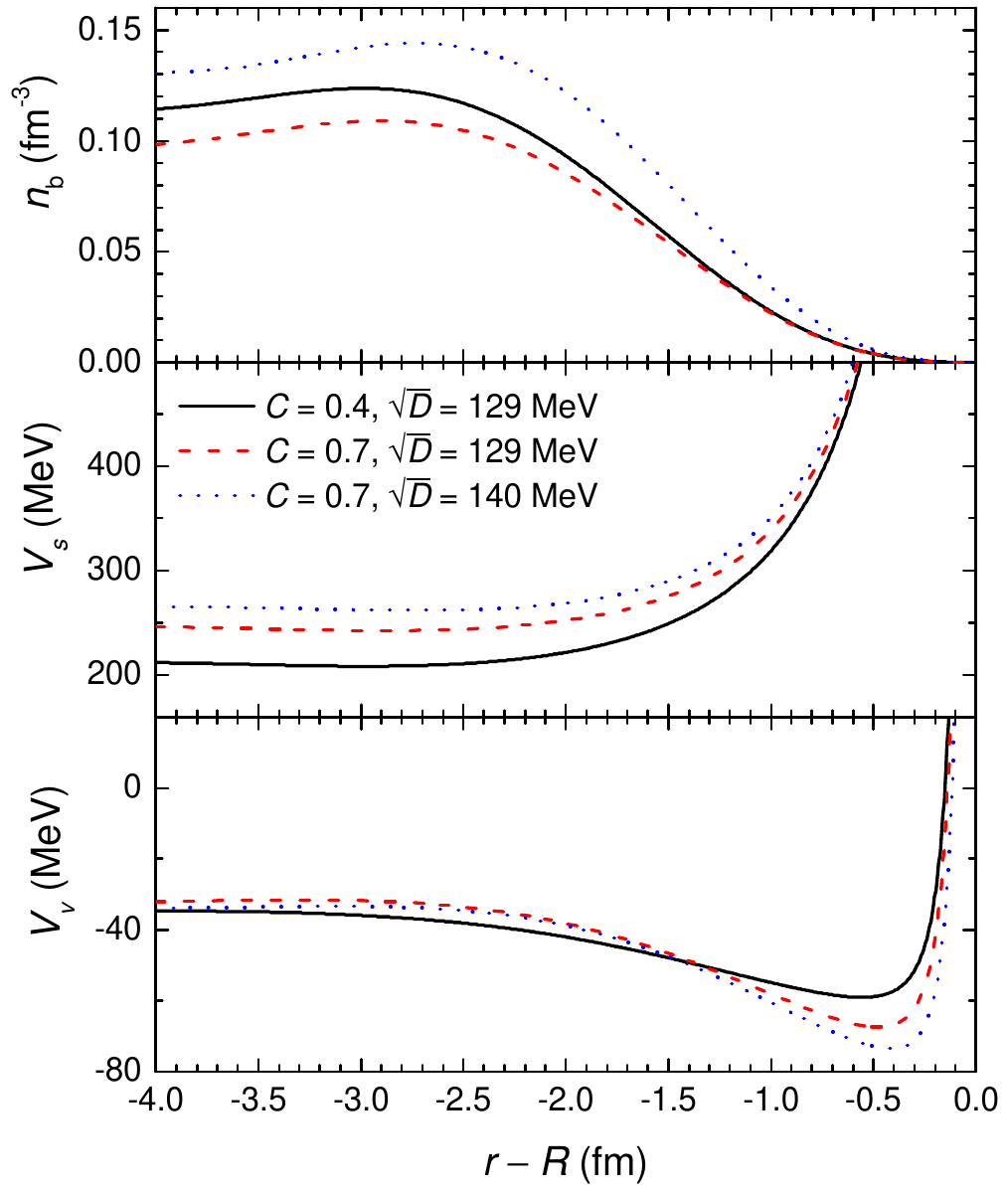}
\caption{\label{Fig:Surf_comp} Density and potential profiles on the surfaces of large strangelets obtained with various parameter sets.}
\end{figure}

In Fig.~\ref{Fig:Surf_comp} we compare the surface structures of strangelets ($A\approx 10^5$) obtained with various parameter sets.
Similar to the cases in Fig.~\ref{Fig:Surf_D129C07}, the density starts to drop at $R-r\approx 2.5$-3 fm and reaches zero at $r=R$.
The internal density of a strangelet increases with $D$ and decreases with $C$, and is close to the bulk density $n_0$ of SQM at $P=0$,
which is indicated in Table~\ref{table:prop}. For larger $n_0$, it is found that the density drops faster on the surface. Aside
from the confinement term in Eq.~(\ref{Eq:mnbC}), increasing the perturbative strength $C$ slightly modifies the surface structure.

\begin{figure}
\includegraphics[width=\linewidth]{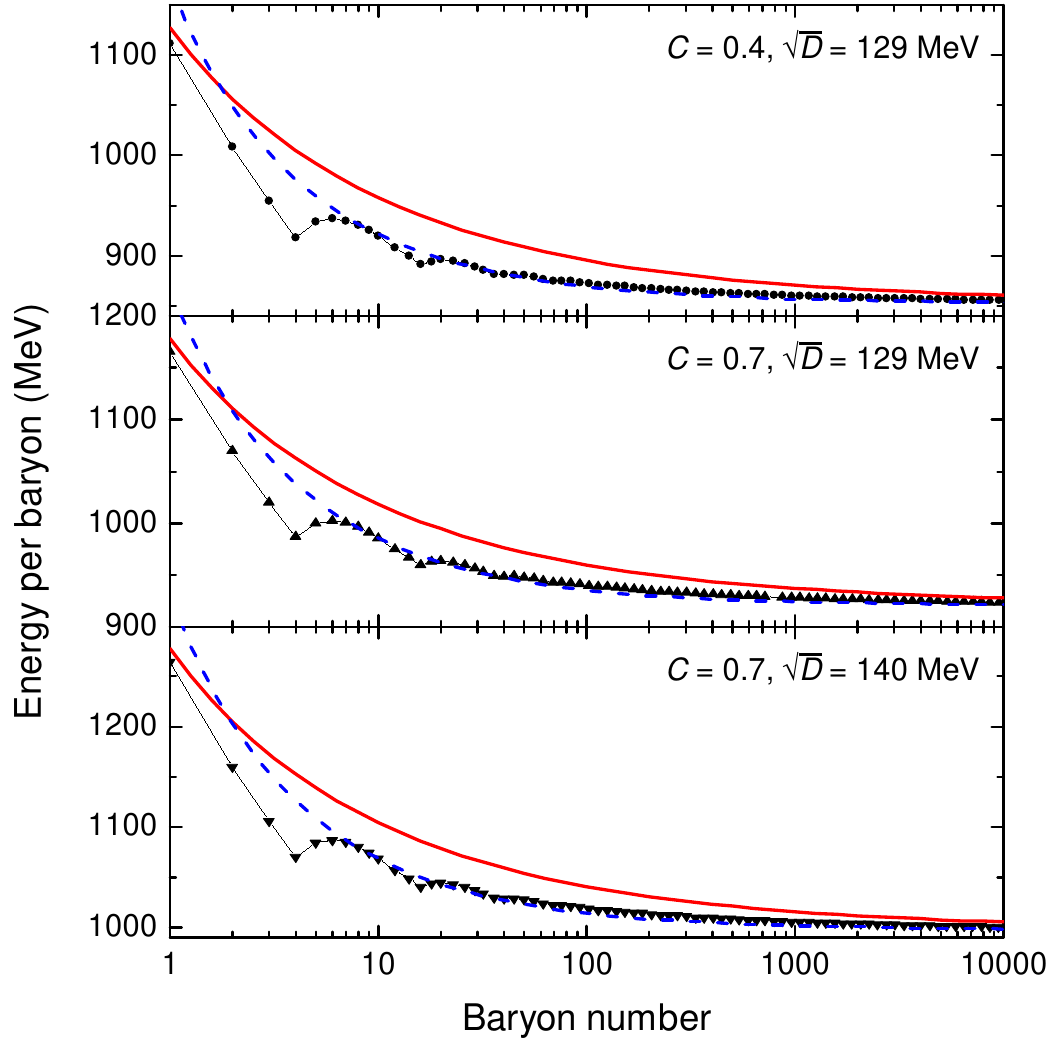}
\caption{\label{Fig:EPA} The energy per baryon of strangelets in MFA (symbols connected with black lines) and
is compared with those obtained with MRE method (red solid curve). A modification to MRE method is introduced
and the results are indicated with the blue dashed curves. Same convention is adopted in the following figures.}
\end{figure}

The energy per baryon of strangelets are presented in Fig.~\ref{Fig:EPA}, which are compared with those obtained with MRE method.
As expected, due to shell effects, strangelets with certain baryon numbers ($A=4$, 16, $\ldots$) are more stable than the
neighboring ones. The corresponding magic numbers for $u$, $d$, $s$ quarks are 6, 24, \ldots, which are exactly 3 times
the magic numbers of finite nuclei~\cite{Sun2018_CPC42-25101}. Since $s$ quarks do not appear until $A>16$, the numbers of $u$ and
$d$ quarks at $A=4$ and 16 are identical to the most stable nuclei in our universe, i.e., ${}^4$He and ${}^{16}$O. However, at larger
$A$, such connections may be altered and varies with the choices of parameters, where the magic numbers (20, 28, 50, 82, \ldots) for
finite nuclei may not appear. Alternatively, the shell structures are not predicted by MRE method since it only contains the average
effects. Meanwhile, the energy per baryon
exhibits an overall trend that decreases with $A$ and approaches to the bulk value in Table~\ref{table:prop}. However, strangelets
obtained with MRE method are more massive, indicating a too large surface tension. In order for strangelets to be realistic
and exist stably without decay, the Witten-Bodmer hypothesis~\cite{Bodmer1971_PRD4-1601, Witten1984_PRD30-272} should be fulfilled,
while the masses of strange stars reach 2 $M_\odot$~\cite{Antoniadis2013_Science340-6131}. In such cases, the parameter set $C = 0.7$
and $\sqrt{D} = 129$ MeV is more reasonable and the corresponding lower limit of baryon number for stable strangelets is
$A_\mathrm{min}\approx 553$.

\begin{table*}
\caption{\label{table:prop} The bulk properties of SQM at zero external pressure and fitted liquid-drop
parameters for the energy per baryon of strangelets in Fig.~\ref{Fig:EPA}. The surface tension and curvature
term corresponding to the fitted parameters are given as well. }
\begin{tabular}{cc|ccc|cccc|cccc|c} \hline \hline
\multicolumn{2}{c|}{Parameters} & \multicolumn{3}{c|}{Bulk properties}
& \multicolumn{4}{c|}{MFA} & \multicolumn{4}{c|}{MRE method}   &  \\ \hline
$C$ & $\sqrt{D}$ &    $n_0$     & ${E_0}/{n_0}$   &  $f_S$ &
$\alpha_S$  & $\alpha_C$   & $\sigma$   &$\lambda$ &	
$\alpha_S$  & $\alpha_C$   & $\sigma$   &$\lambda$ & $\sigma^\mathrm{MFA}/\sigma^\mathrm{MRE}$\\
 & MeV & fm${}^{-3}$ & MeV & & MeV & MeV & MeV/fm${}^2$ & MeV/fm & MeV & MeV & MeV/fm${}^2$ & MeV/fm & \\ \hline
0.4 & 129 & 0.11  & 850.91 & 0.20  & 56 & 177 & 2.7 & 5.49 & 190.5 & 86.1 & 9.247 & 2.67 & 0.29 \\
0.7 & 129 & 0.099 & 918.94 & 0.056 & 54 & 172 & 2.4 & 5.12 & 173.5 & 85.7 & 7.681 & 2.54 & 0.31 \\
0.7 & 140 & 0.13  & 995.77 & 0.14  & 61 & 185 & 3.3 & 6.03 & 191.1 & 90.9 & 10.18 & 2.96 & 0.32 \\
\hline
\end{tabular}
\end{table*}

Based on Fig.~\ref{Fig:EPA}, the surface tension and curvature term of SQM can be extracted with a liquid-drop type
formula~\cite{Oertel2008_PRD77-074015}
\begin{equation}
\frac{M}{A} = \frac{E_0}{n_0} + \frac{\alpha_S}{A^{1/3}} + \frac{\alpha_C}{A^{2/3}}, \label{Eq:M_ld}
\end{equation}
where $E_0$ is the energy density of SQM at $P=0$ and $E_0/n_0$ corresponds to the minimum energy per baryon indicated in
Table~\ref{table:prop}. By fitting to the data in Fig.~\ref{Fig:EPA} with Eq.~(\ref{Eq:M_ld}), one can obtain the surface
tension $\sigma$ and curvature term $\lambda$ with the fitted parameter $\alpha_S$ and $\alpha_C$, which are given by
\begin{eqnarray}
\sigma  &=& \alpha_S \left(\frac{ n_0^2}{36 \pi}\right)^{1/3}, \label{Eq:sigma} \\
\lambda &=& \alpha_C \left(\frac{n_0}{384 \pi^2}\right)^{1/3}. \label{Eq:lambda}
\end{eqnarray}
The obtained results are presented in Table~\ref{table:prop}. With the fitted parameters, Eq.~(\ref{Eq:M_ld}) can well reproduce
the energy per baryon in Fig.~\ref{Fig:EPA}. By examine the dependence of the surface tension $\sigma$ and curvature term $\lambda$
on the parameters $C$ and $D$, it is found that the linear confinement increases $\sigma$ and $\lambda$ while the perturbative
interaction does the opposite. If SQM is absolutely stable, according to pulsar observations~\cite{Xia2014_PRD89-105027,
Xia2015_CAA56-79, Peng2018_JPSCP20-011022}, the parameter set $C=0.7$ and $\sqrt{D} = 129$ MeV is more reasonable, which gives
$\sigma \approx 2.4$ MeV/fm${}^2$ and $\lambda \approx 5.12$ MeV/fm. For unstable SQM, slightly larger values are obtained
for $\sigma$ and $\lambda$. The linear dependences of $\sigma$ and $\lambda$ on the saturation density $n_0$ of SQM are observed,
which gives $\sigma \approx 23.5 n_0$ and $\lambda \approx 1.8 + 32.8 n_0$ with the units corresponding to
those in Table~\ref{table:prop}. Meanwhile, it is found that the MRE method overestimates the surface tension and underestimates
the curvature term. Note that MRE method was initially proposed by reproducing bag model results~\cite{Madsen1994_PRD50-3328},
where quark confinement is reached with an infinite wall. Thus we shall not expect MRE method to reproduce our results
since confinement is attained in a different mechanism. As indicated in Fig.~\ref{Fig:Surf_D129C07} and Fig.~\ref{Fig:Surf_comp},
the densities slowly approach to zero on the quark-vacuum interface as quark masses approach to infinity according to
Eq.~(\ref{Eq:mnbC}), which correspond to linear confinement. For MRE method to roughly reproduce our results obtained with
equivparticle model, as was done in Ref.~\cite{Xia2017_JPCS861-012022}, we take the surface strength factor $\eta_\mathrm{s}
\approx 0.3$ in Eq.~(\ref{Eq:fs}) and curvature strength factor $\eta_\mathrm{c}\approx 0.1$ in Eq.~(\ref{Eq:fc}).
The results obtained with the modified MRE method are then presented in Fig.~\ref{Fig:EPA} with blue dashed curves, which
coincide with MFA at $A\gtrsim 8$.

\begin{figure}
\includegraphics[width=\linewidth]{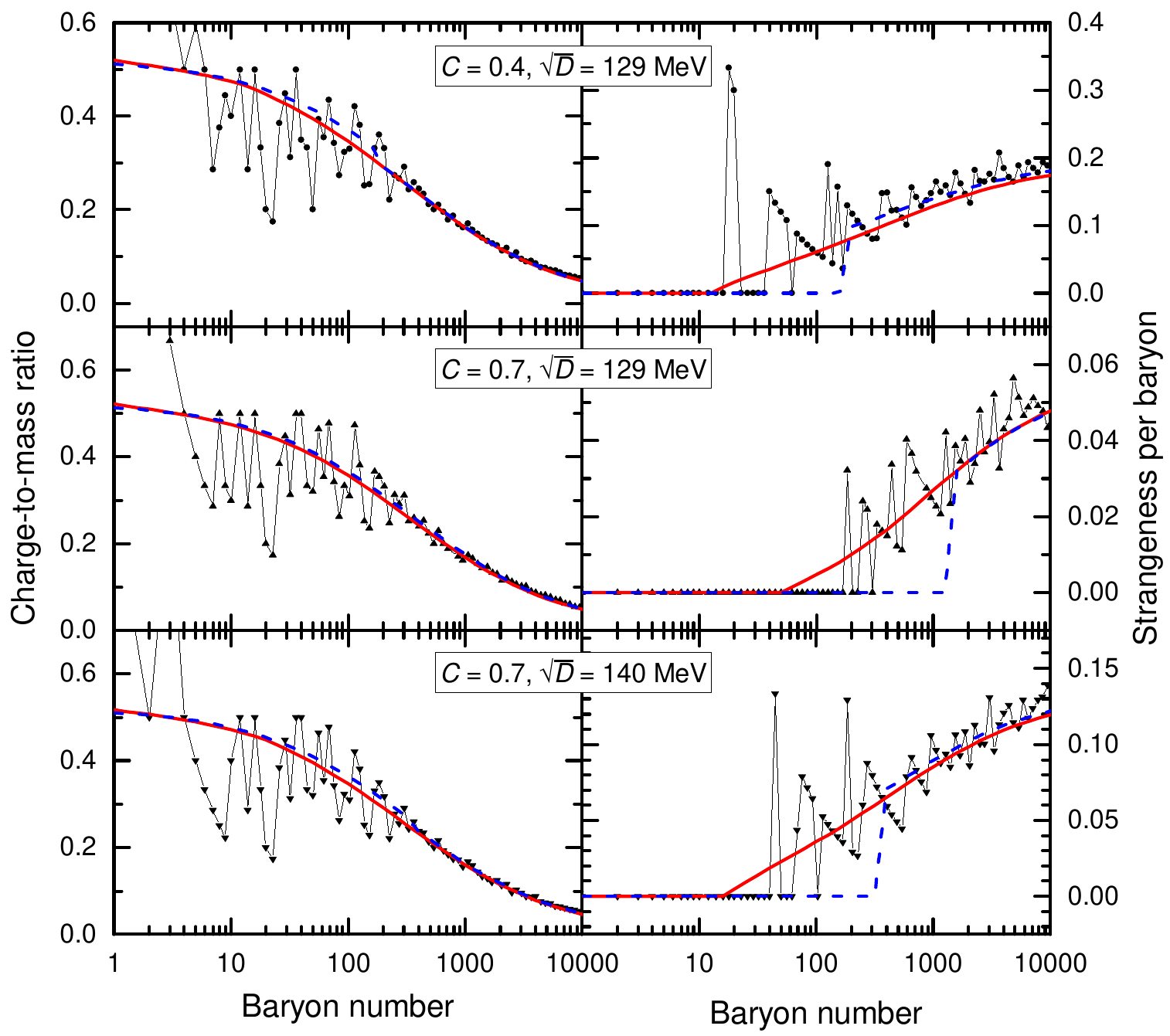}
\caption{\label{Fig:fzs} Left: Charge-to-mass ratio ($f_Z = Z/A$) of $\beta$-stable strangelets;
Right: Strangeness per baryon ($f_S = S/A$) of strangelets obtained in various methods.}
\end{figure}

The charge-to-mass ratio $f_Z$ and strangeness per baryon $f_S$ of $\beta$-stable strangelets are presented in
Fig.~\ref{Fig:fzs}. Despite the large differences on their masses, it is found that MRE method reproduces the average
values of $f_Z$ and $f_S$. Due to shell effects, strangelets with $S=6$, 24, 54, 60, $\ldots$ are more stable than
others, where the strangeness per baryon $f_S = S/A$ is decreasing with $A$. The overall trends are observed for
$f_Z$ and $f_S$ as functions of $A$, where $f_Z$ reduces to 0 and $f_S$ increases to the bulk value in
Table~\ref{table:prop} as $A\rightarrow \infty$. For larger bulk values of $f_S$, the onset baryon number
becomes smaller, which varies from $A\approx 20$ to 200.

\begin{figure}
\includegraphics[width=\linewidth]{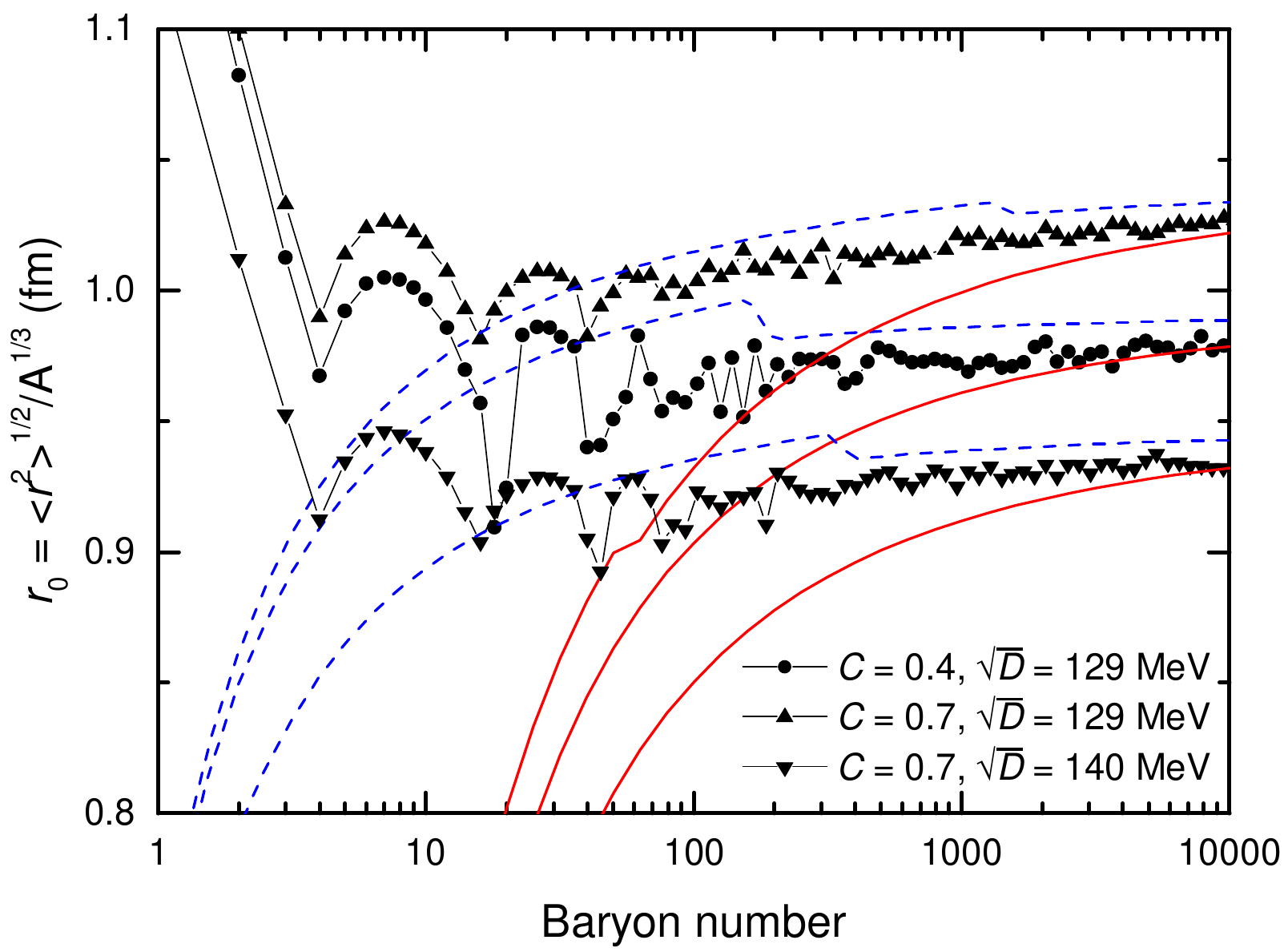}
\caption{\label{Fig:r0} The ratio of root-mean-square radius to baryon number for strangelets obtained in various methods.}
\end{figure}

To show the compactness of strangelets, in Fig.~\ref{Fig:r0} we present the ratio of root-mean-square radius to baryon
number $r_0$ for $\beta$-stable strangelets, which is compared with those obtained with MRE method. The root-mean-square
radius of a strangelet can be determined by
\begin{equation}
\langle r^2 \rangle = \frac{\int 4\pi r^4 n_\mathrm{b}(r) \mbox{d}r}{A}. \label{Eq:rms}
\end{equation}
Based on the density profiles inside a strangelet, the values of $r_0$ can then be determined by $r_0 = \langle r^2 \rangle^{1/2}/A^{1/3}$.
When MRE method is adopted, the density distribution $n_\mathrm{b}(r)$ can be approximated by 2 parts, i.e.,
\begin{equation}
n_\mathrm{b}(r) = \sum_{i=u, d, s} \frac{\nu_i^3}{3 \pi^2} + \frac{A^\mathrm{surf}}{4 \pi R^2} \delta(r-R),
\label{Eq:nb_MRE}
\end{equation}
where the first part corresponds to the volume term and the second part is the surface term with $A^\mathrm{surf} = A
- \sum_{i=u, d, s}{4\nu_i^3 R^3}/{3\pi}$. For larger strangelets, $r_0$ obtained with different methods approach to a
same value, which is related to $n_0$ in Table~\ref{table:prop}. As the baryon number $A$ decreases, $r_0$ remains
almost constant until a sudden increase is observed at $A\lesssim 6$, while $r_0$ predicted by MRE method is decreasing.
Note that the density distribution in Eq.~(\ref{Eq:nb_MRE}) is not valid for small strangelets, where the quark depletion
is assumed to take place only on the surface at $r=R$. With the modification to MRE method, the obtained $r_0$
coincides with those given by MFA at $A \gtrsim 10$.

\section{\label{sec:con}Conclusion}
We study the interface effects of SQM in equivparticle model, where both linear confinement and leading-order perturbative
interactions are included with density-dependent quark masses. In mean-field approximation, the properties of strangelets
are presented, and compared with those of the MRE method. The surface tension and curvature term due to the
quark-vacuum interface on the surface of a strangelet is then investigated. By increasing the confinement strength, it is
found that the surface tension and curvature term of SQM become larger, while the perturbative interaction does the opposite.
For those parameters constrained according to the 2$M_\odot$ strange star, the surface tension is $\sim$2.4 MeV/fm${}^2$,
while unstable SQM indicates a slightly larger surface tension. Since MRE method was initially proposed to reproduce bag
model results with quarks confined in an infinite well, one should not expect that the original MRE method to reproduce the
results obtained in the equivparticle model, where linear confinement is also considered. It is found that the original
MRE method overestimates the surface tension and underestimate the curvature term. For MRE method to roughly reproduce
the results in MFA, we introduced the surface and curvature strength factors, which are respectively about 0.3 and 0.1.

Finally, it should be pointed out that there are many factors affecting the interface effects of SQM, which are
not involved in the present study. For example, the effect of color superconductivity should play an important
role~\cite{Alford2001_PRD64-074017, Madsen2001_PRL87-172003, Oertel2008_PRD77-074015}. Our calculation is performed
at zero temperature, i.e., the temperature effect~\cite{Ke2014_PRD89-074041, Gao2016_PRD94-094030} was not considered.
Therefore, further studies are necessary.

\section*{ACKNOWLEDGMENTS}
This work was supported by National Natural Science Foundation of China (Grant Nos.~11705163, 11621131001, 11575201, 11575190,
11525524, 11505157, and 11475110), the Physics Research and Development Program of Zhengzhou University (Grant No.~32410017),
and the U.S. National Science Foundation (Grant No. PHY 1608959). The computation for this work was supported by the
HPC Cluster of SKLTP/ITP-CAS and the Supercomputing Center, CNIC, of the CAS.

%

\end{document}